\documentclass[12pt,a4paper]{article}
\usepackage{amsmath}
\usepackage{graphicx}

\begin{document}

\title{%
  Higgs Search in the $WW^*$ Decay Mode at  
  Photon Linear Colliders}

\author{%
  E.~Boos$^{1,2}$,
  V.~Ilyin$^2$,
  D.~Kovalenko$^2$,
  T.~Ohl$^1$,\\
  A.~Pukhov$^2$,
  M.~Sachwitz$^3$, 
  H.J.~Schreiber$^3$\\
  \hfil\\
  $^1$Institut f\"ur Kernphysik, TU Darmstadt, Germany \\
  $^2$Institute of Nuclear Physics, \\
      Moscow State University, 119899, Moscow, Russia \\
  $^3$DESY-Zeuthen, Germany}

\date{%
  DESY 98-004\\
  IKDA 98/2\\
  hep-ph/9801359\\
  January 1998\\
  \hfil}

%%% \preprintno{%
%%%   DESY 98-004\\
%%%   IKDA 98/2\\
%%%   hep-ph/9801359}
%%% \date{January 1998}

%%%%%%%%%%%%%%%%%%%%%%%%%%%%%%%%%%%%%%%%%%%%%%%%%%%%%%%%%%%%%%%%%%%%%%%%
\maketitle
\begin{abstract}
  We present the results of calculations for the process
  $\gamma\gamma\to W+2\mathop{\textrm{fermions}}$ at a future Photon
  Linear Collider~(PLC).  The calculations include at the same time
  the next-to-leading order Higgs signal and the complete tree level
  gauge boson background.  We present numerical results in the
  intermediate mass Higgs region $140 \mathop{\textrm{GeV}}< M_H < 2M_W$.
  We propose strategies for the determination of Higgs properties
  using the leptonic and hadronic final states for both polarized and
  unpolarized photon beams.
\end{abstract}

%%%%%%%%%%%%%%%%%%%%%%%%%%%%%%%%%%%%%%%%%%%%%%%%%%%%%%%%%%%%%%%%%%%%%%%%
\section{Introduction}
\label{ref:intro}

The search for a Higgs boson, the last unobserved particle predicted
by the minimal electro-weak Standard Model~(SM), is a primary
objective for experiments at operating colliders, as well as for the
planning of future colliders.  The potential discovery of a
fundamental Higgs particle--either at a hadron collider (Tevatron,
LHC) or at a $e^+e^-$-collider (LEP2, Next Linear Collider)--will
initiate detailed studies and measurements of its properties like
spin, parity, decay branching fractions, total width, and the couplings
with gauge bosons and fermions.  Any deviation of these results from
the SM predictions will be a signature of new physics beyond the SM.

Electro-weak precision tests at LEP1 are sensitive to the symmetry
breaking sector and a fundamental Higgs through radiative corrections
to the electro-weak observables.  The minimal SM and the most popular
extensions (supersymmetry) favour a neutral Higgs boson in the
intermediate-mass region between~$M_Z$ and just above~$2M_Z$.  In this
mass range, the search for the Higgs boson and the study of its
properties are delicate phenomenological and experimental problems,
mostly because of large irreducible backgrounds.

In fact, the complementary information provided by different colliders
has to be used~\cite{zerwas,gunion1} and a Photon Linear Collider~(PLC)
hosted at a $e^+e^-$ Linear Collider~\cite{telnov} plays an important
r\^ole in this context.  For example, the total Higgs
width~$\Gamma_H^{\text{tot}}$, which is related directly to the
fundamental couplings, can be determined in a model-independent way by
simultaneously using results from various machines~\cite{gunion2}.  This
procedure combines the branching ratios $\mathrm{BR}(H^0\to b\bar b)$
and $\mathrm{BR}(H^0\to\gamma\gamma)$, measured at an $e^+e^-$ or
hadron machine, with the width $\Gamma(H^0\to\gamma\gamma)$
measured at the PLC.  The detailed study of the $\gamma\gamma H^0$
vertex at the required precision is only possible at the PLC.

In the intermediate Higgs mass range
$140\mathop{\textrm{GeV}}<M_H<2M_Z$, the dominant decay mode is into
$WW$ pairs.  This final state is contaminated by a huge tree level
gauge boson background and it is generally expected that it will be
very difficult to use for Higgs physics at a PLC~\cite{bbc}.

Recently, this question has been reanalysed~\cite{ginzburg} with the
conclusion that both the $WW^*$ and $WW$ modes can be used if a
reasonable (on the order of 10\,GeV) resolution for the $WW$ invariant
mass can be achieved experimentally.  However, the calculations
in~\cite{ginzburg} are based on on-shell matrix elements continued
from above the $WW$ pair production threshold.  This method gives the
correct signal to background~(S/B) ratio above the threshold, of
course.  However, as will be shown below, it overestimates the S/B
ratio significantly below the threshold.  Indeed, it has already been
pointed out~\cite{boos-ohl} that the correct background rate below the
$WW$ threshold is dominated by the contribution of nonresonant diagrams.
In the present paper we investigate the possibilities for searching
for the Higgs boson in that mass range from about 140\,GeV to $2M_W$
at a PLC in more detail.

After a short discussion of Higgs production at a PLC in
section~\ref{sec:Higgs@PLC}, we present in this paper an analysis for
the Higgs signal process including all important corrections
(cf.~section~\ref{sec:calc}) 
\begin{subequations}
\begin{align}
\label{eq:main}
  \gamma \gamma &\to H^0 \to W + 2\mathop{\textrm{fermions}}\\
\intertext{%
  together with the complete tree level calculation of the
  corresponding background process}
\label{eq:mainbg}
  \gamma \gamma &\to W + 2\mathop{\textrm{fermions}}
\end{align}
\end{subequations}
which includes the subsequent decay of the $W^*$ into quark or lepton
pairs as well as all other SM contributions. In the remainder of the
paper, we will often write $\gamma\gamma\to WW^*$ for brevity.  In
section~\ref{sec:xsect} we present numerical results for these cross
sections and we investigate two extreme scenarios for the operation of
a PLC in section~\ref{sec:Scenarios@PLC}.  We conclude in
section~\ref{sec:concl}.

%%%%%%%%%%%%%%%%%%%%%%%%%%%%%%%%%%%%%%%%%%%%%%%%%%%%%%%%%%%%%%%%%%%%%%%%
\section{Higgs Searches at a Photon Linear Collider}
\label{sec:Higgs@PLC}

The optimal strategy for the Higgs search and the experimental
determination of its properties at any collider depends strongly on
the value of the Higgs mass under consideration. At the PLC we can
distinguish three mass ranges for the search for the intermediate mass
Higgs that require different approaches.

\subsection{Light Higgs: $M_H<140\mathop{\textmd{\textrm{GeV}}}$}
In the mass range below 140\,GeV, a Higgs boson decays predominantly
into~$b\bar b$ and the Higgs signal at a PLC can be extracted from the
background using highly polarized photon beams (cf.~one of the latest
studies~\cite{fadin} where the two loop QCD corrections have been
taken into account).  For collider energies up to about
300\,--\,400\,GeV, the background from resolved photon processed
remains small enough~\cite{eboli}.  The potentially large four-fermion
background like $\gamma\gamma\to e^+e^-b\bar b$ where the $e^+e^-$
pair is lost in the beam pipe can be suppressed by a proper cut using
the different kinematics of the signal~\cite{boos1}.  In this case
highly polarized photon beams are not sufficient for rejecting the
background, because the initial photon polarization is shared among
the polarizations of the four final state fermions and therefore no
strong suppression of $b\bar b$ production occurs.

\subsection{Heavy Higgs: $M_H>2M_Z$}
In the mass range above $2M_Z$, the dominant Higgs decay mode into
$W$-pairs is obscured by the huge tree level background
$\gamma\gamma\to WW$~\cite{bbc}.  It is possible to utilize the
interference of the Higgs signal with the $WW$
background~\cite{truong}, which manifests itself as a resonant dip in
the $WW$ invariant mass distribution for Higgs studies.  However, in
this case a more detailed simulation of the four-fermion final state,
including all tree level diagrams is mandatory.  Alternatively, one
can use the $ZZ$ mode with semileptonic four-fermion final states
consisting of one neutral lepton pair and two jets~\cite{bbc, haber}.
But the one-loop $\gamma\gamma\to ZZ$ background becomes important
when the Higgs mass exceeds about 350\,--\,400\,GeV
~\cite{jikia,berger} and makes a search for the Higgs even more
problematic in this region.

\subsection{%
  Intermediate Mass Higgs: $140\mathop{\textmd{\textrm{GeV}}}<M_H<2M_Z$}
In the remaining region for the Higgs mass from 140\,GeV to $2M_Z$,
the determination of the Higgs properties is a challenging
experimental task.  Here, the decay mode of Higgs to $WW$ dominates as
well, but it is plagued by a huge tree level background.
Nevertheless, it has been shown in~\cite{ginzburg} that a signal can
be extracted for Higgs masses above the $W$ pair production threshold.

In the remainder of this paper, we will discuss the $WW^*$ final state
for Higgs masses below the $WW$ threshold in more detail.  Despite the
fact that the S/B ratio has been overestimated significantly
in~\cite{ginzburg}, we will present strategies for measuring Higgs
properties in this region as well.

%%%%%%%%%%%%%%%%%%%%%%%%%%%%%%%%%%%%%%%%%%%%%%%%%%%%%%%%%%%%%%%%%%%%%%%%
\section{%
  Cross Section and Background for
  $\gamma\gamma\to\bar\nu\mu^-W^+$ and $\gamma\gamma\to\bar udW^+$}
\label{sec:xsect}

%%%%%%%%%%%%%%%%%%%%%%%%%%%%%%%%%%%%%%%%%%%%%%%%%%%%%%%%%%%%%%%%%%%%%%%%
\subsection{Calculational Procedures}
\label{sec:calc}

\begin{figure}
  \begin{center}
    \includegraphics[bb=100 350 500 680,width=12cm]{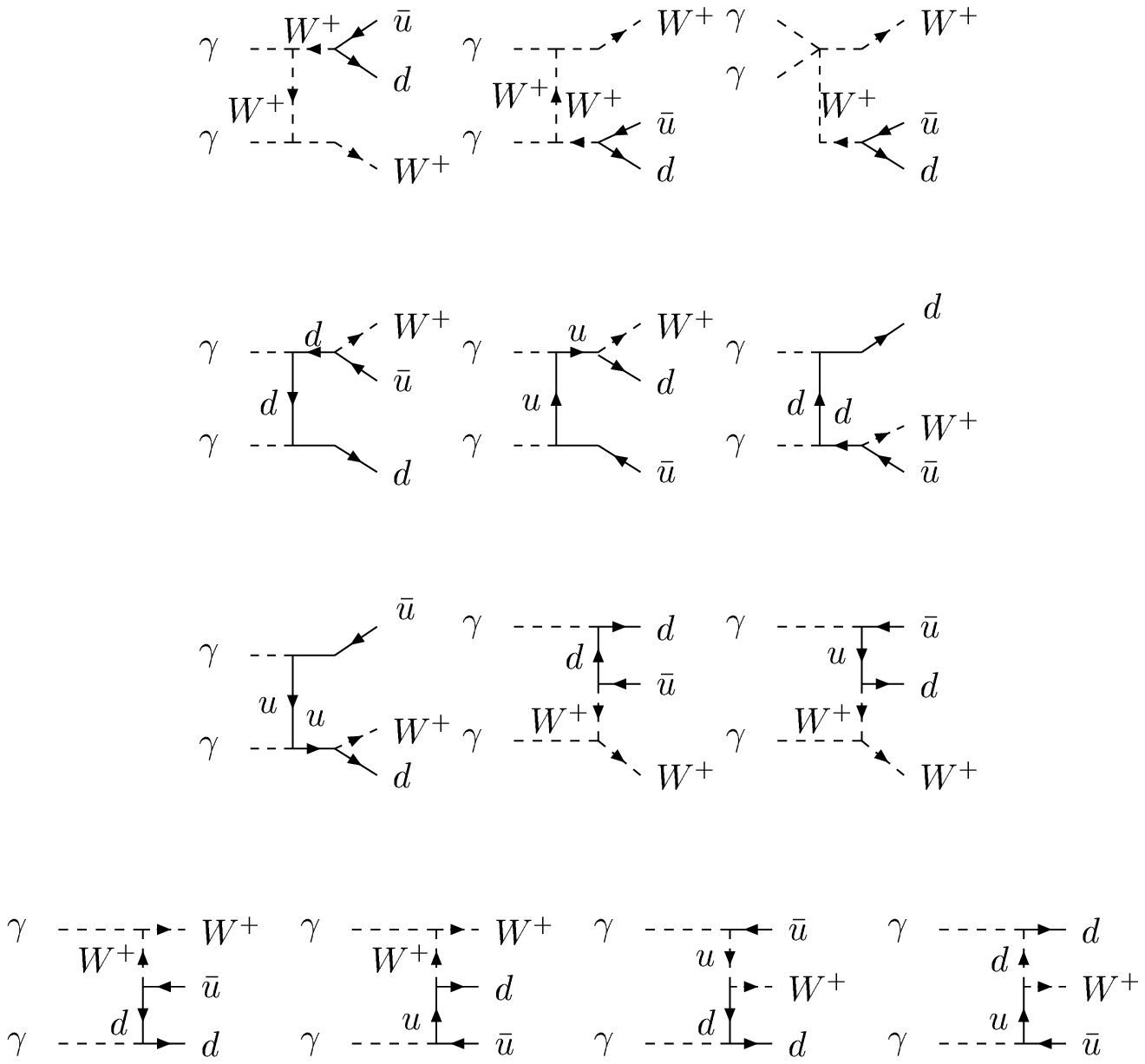}
  \end{center}
  \caption{\label{fig:udW}%
    Feynman diagrams for $\gamma\gamma\to\bar udW^+$.}
\end{figure}

\begin{figure}
  \begin{center}
    \includegraphics[bb=100 550 500 630,width=12cm]{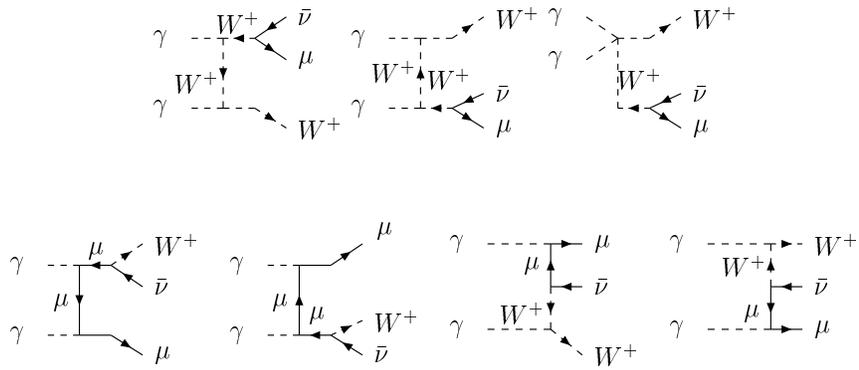}
  \end{center}
  \caption{\label{fig:numuW}%
    Feynman diagrams for $\gamma\gamma\to\bar\nu_\mu\mu^-W^+$.}
\end{figure}

The SM tree-level diagrams contributing to the background process
$\gamma\gamma\to WW^*$ for the quark pair final state $\bar udW^+$ are
shown in figure~\ref{fig:udW} and for the semileptonic final state
$\bar\nu\mu^-W^+$ in figure~\ref{fig:numuW}.  Only hadronic decays of
the $W$ will be selected to allow for invariant mass cuts.

The calculations were performed using the software package
\texttt{CompHEP} Version~3.2~\cite{comphep} for unpolarized and
polarized photon beams.  In the polarized case, the unpolarized
initial state matrix~$\sum\epsilon^\mu\epsilon^{*\nu}=-g^{\mu\nu}$ in
Feynman gauge is replaced by the axial gauge expressions
\begin{equation}
\label{eq:matrix}
  T^{\mu\nu}_\lambda
    = \frac{1}{2} \left[ - g^{\mu\nu}
                         + \frac{p^\mu p^{\prime\nu}
                                  + p^{\prime\mu} p^\nu}{pp'}
                         + i\lambda\frac{\epsilon^{\mu\nu\rho\sigma}
                                          p_\rho p'_\sigma}{pp'}
                  \right]
\end{equation}
where $p$ and~$p'$ are the light-like Lorentz momenta of the incoming
photons with $pp'= s_{\gamma\gamma}/2$ and $\lambda=\pm1$ is the
helicity, with the convention $\epsilon^{0123}=+1$.

In order to calculate the Higgs signal contribution, the effective
$H\gamma\gamma$ vertex has been implemented in \texttt{CompHEP}.  In
addition, all important corrections have been included: NLO and NNLO
QCD corrections to the running quark masses~\cite{MQrunning} as well
as the three particle partial Higgs decay width $\Gamma_{H\to WW^*}$
with $W^*\to 2\mathop{\textrm{fermions}}$~\cite{WWstar}.  The accuracy
of the implementation of the $H\gamma\gamma$ vertex has been verified
by finding very good agreement with the results from the program
\texttt{HDECAY}~\cite{HDECAY}.

%%%%%%%%%%%%%%%%%%%%%%%%%%%%%%%%%%%%%%%%%%%%%%%%%%%%%%%%%%%%%%%%%%%%%%%%
\subsection{Background Cross Section}
\label{sec:results}

\begin{figure}
  \begin{center}
    \includegraphics[width=12cm]{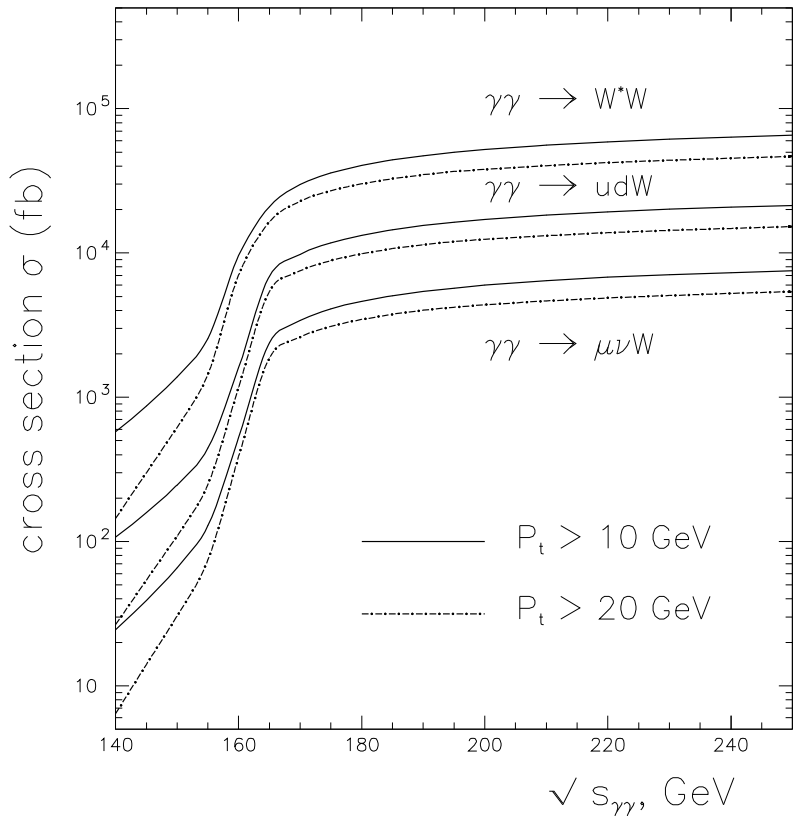}
  \end{center}
  \caption{\label{fig:sigma_all}%
    Cross section for the process $\gamma\gamma\to WW^*$.}
\end{figure}

Figure~\ref{fig:sigma_all} shows the background cross section
$\gamma\gamma\to WW^*$ together with the leptonic and quark subsets
for $p_\perp$-cuts of 10 and 20\,GeV on both fermions as a function of
the photon-photon energy $\sqrt{s_{\gamma \gamma}}$.
As expected, the cross sections vary
strongly with the energy $\sqrt{s_{\gamma\gamma}}$ in the threshold
region from 140\,GeV to $2M_W$.  Therefore, a careful consideration of
the background and signal rates is mandatory.

%%%%%%%%%%%%%%%%%%%%%%%%%%%%%%%%%%%%%%%%%%%%%%%%%%%%%%%%%%%%%%%%%%%%%%%%
\section{%
  Scenarios for the Higgs Search in the $WW^*$ Mode
  at a PLC below the $WW$ Threshold}
\label{sec:Scenarios@PLC}

Depending on the advances in technology, it will be possible to
operate a future PLC in a variety of modes (see~\cite{telnov} for a
recent review).  In order to cover a wide range of options, we
consider two extreme scenarios: scenario~I is a more conservative
\emph{wide-band} PLC and scenario~II is a technically more demanding
\emph{narrow-band} PLC.

\begin{itemize}

  \item In the technically most conservative scenario~I,
    laser backscattering is used to turn a first stage $e^+e^-$
    Next Linear Collider operating at the $t\bar t$ threshold 360\,GeV
    into a PLC with colliding $\gamma\gamma$ beams.
    A rather broad photon energy spectrum with
    a maximum energy at $\approx80\%$ of the incident electron energy
    is expected for the colliding photons, which will be unpolarized.

  \item In the technically more challenging scenario~II, the
    energy of the electron beam, the frequency of the laser, the
    respective polarizations and the geometry of the conversion and
    interaction region are tuned to produce highly polarized
    colliding $\gamma\gamma$ beams with energy and luminosity
    comparable to the underlying $e^+e^-$ machine~\cite{telnov}.
    The accelerator parameters can be chosen so that the differential
    $\gamma\gamma$ luminosity has a peak at a particular energy
    (e.\,g.~at $\sqrt{s_{\gamma\gamma}}=M_H$) with a width of
    approximately 15\%~\cite{telnov}.

    Until realistic energy spectra for the photons can be made
    available by accelerator physicists, we use a simple step function
    of appropriate width for our simulations.  We do not expect our
    conclusions to depend sensitively on the shape of the spectrum.
    In a real analysis, a spectrum determined experimentally from a
    reference process will be used.

\end{itemize}

Obviously, scenario~II is best suited for studying Higgs properties,
provided that the Higgs mass will be known at the start of the
operation of a PLC.  In fact, if Nature has chosen an intermediate
mass Higgs for electro-weak symmetry breaking, it is likely that its
mass will have been determined by $e^+e^-$- or hadron-machines at the
time a PLC can be commissioned.  Nevertheless, we will consider the
wide-band scenario~I as well, in order to show what could be achieved
with a technically more conservative design and in order to gauge the
gain in physics results provided by scenario~II.

%%%%%%%%%%%%%%%%%%%%%%%%%%%%%%%%%%%%%%%%%%%%%%%%%%%%%%%%%%%%%%%%%%%%%%%%
\subsection{Scenario~I: Wide-Band PLC}
\label{sec:wide-band}

\begin{table}
  \begin{center}
    \setlength{\arraycolsep}{1em}
    \renewcommand{\arraystretch}{1.4}
    \begin{tabular}{c|c|c}
                        & $\gamma\gamma\to H^0\to\bar\nu\mu^-W^+$
                        & $\gamma\gamma\to\bar\nu\mu^-W^+$  \\
                        & signal/fb   & BG/fb               \\\hline
       no $p_\perp$-cut & 1.64        &  2752               \\
       $p_\perp>10\mathop{\textrm{GeV}}$
                        & 1.26        &  2464               \\
       $p_\perp>20\mathop{\textrm{GeV}}$
                        & 0.43        &  1779
    \end{tabular}
  \end{center}
  \caption{\label{tab:wide-semilep}%
    Cross sections for $\gamma\gamma\to\bar\nu\mu^-W^+$ with
    different $p_\perp$-cuts, assuming a Compton backscattering
    $\gamma$-spectrum (scenario~I).}
\end{table}

Considering the semileptonic case first, we obtain in scenario~I, with
a broad $\gamma$ spectrum up to
$\sqrt{s_{\gamma\gamma}}=0.8\times360\mathop{\textrm{GeV}}$, the cross
sections in table~\ref{tab:wide-semilep}, for a Higgs mass of
$M_H=140\mathop{\textrm{GeV}}$.  The cross sections for the Higgs
signal $\gamma\gamma\to H^0\to\bar\nu\mu^-W^+$ and the background
$\gamma\gamma\to\bar\nu\mu^-W^+$ are given for various $p_\perp$-cuts.
We have neglected the additional irreducible
four-fermion background from the hadronic decay of the $W$ boson.  In
the semileptonic channel, this decreases the S/B ratio only marginally
compared to the huge background coming mostly from the invariant mass
region above the $WW$ threshold.  This region can not be cut out,
because the indeterminate photon energy and the ``lost'' neutrino
conspire to make a full kinematical reconstruction impossible.

\begin{figure}
  \begin{center}
    \includegraphics[width=12cm]{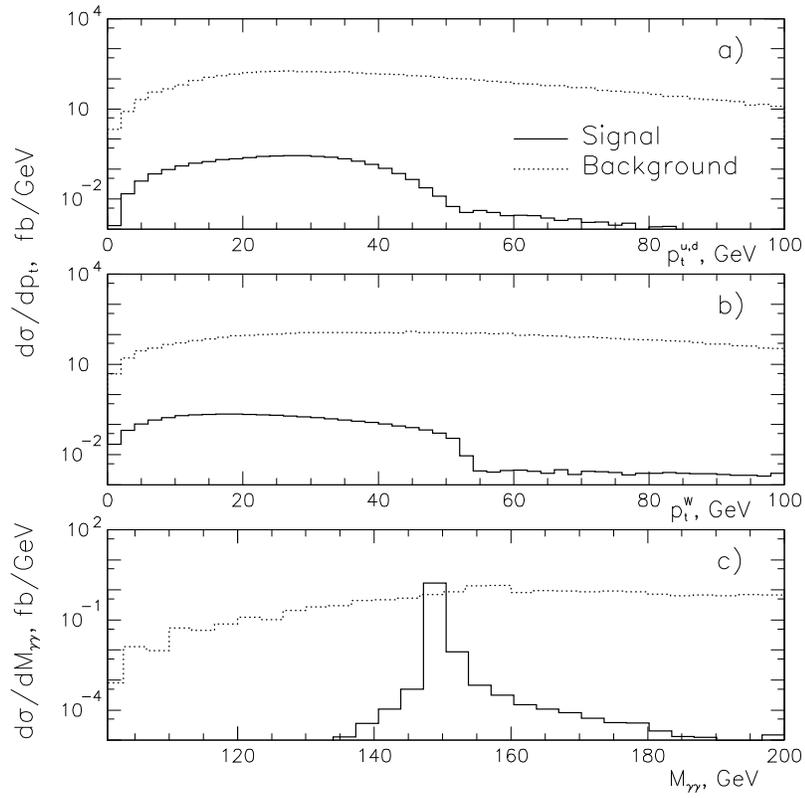}
  \end{center}
  \caption{\label{fig:diff-udw-compt150}%
    Differential cross section for $\gamma\gamma\to\bar udW^+$.  Note
    that the differential distributions are identical for
    $\gamma\gamma\to\bar\nu_\mu\mu^-W^+$, only the normalization
    changes.}
\end{figure}

In table~\ref{tab:wide-semilep}, we observe that the S/B ratio is
prohibitively small, irrespective of the $p_\perp$-cut on the~$\mu$
and on the missing transversal momentum.
An additional cut, $p_\perp<50\mathop{\textrm{GeV}}$, on the
transversal momenta of~$\mu$ and~$W$ and on the missing transversal
momentum improves the S/B ratio by a factor of two only (see
figure~\ref{fig:diff-udw-compt150} and the discussion of the hadronic
final states below), which is insufficient for establishing a signal.
Therefore, we have to conclude that it appears to be hopeless to find
the Higgs signal in the semileptonic final state for scenario~I.

\begin{table}
  \begin{center}
    \setlength{\arraycolsep}{1em}
    \renewcommand{\arraystretch}{1.4}
    \begin{tabular}{c|rr|rr|rr}
       & \multicolumn{2}{c|}{$M_H=140\mathop{\textrm{GeV}}$}
       & \multicolumn{2}{c|}{$M_H=150\mathop{\textrm{GeV}}$}
       & \multicolumn{2}{c}{$M_H=160\mathop{\textrm{GeV}}$}   \\
         cuts  & signal/fb & BG/fb  & signal/fb & BG/fb  & signal/fb & BG/fb
           \\\hline
     canonical &  4.4      & 7286.0 &  7.6      & 7286.0 & 13.6      & 7286.0\\
     optimized &  3.4      &    3.9 &  6.4      &    8.1 & 12.1      &   94.4
    \end{tabular}
  \end{center}
  \caption{\label{tab:wide-hadr}%
    Cross sections for $\gamma\gamma\to\bar udW^+$, assuming a Compton
    backscattering $\gamma$-spectrum (scenario~I).  See the text for a
    description of canonical and optimized cuts.}
\end{table}

Turning our attention to the hadronic final state $\gamma\gamma\to\bar
udW^+$, we obtain the cross sections shown in
table~\ref{tab:wide-hadr} for three different Higgs masses.  If only
the canonical cuts from the Linear Collider studies~\cite{zerwas} are
used, the S/B ratio is as prohibitively small as in the semileptonic
case.  These cuts correspond to the most inclusive coverage of a
realistic detector: a minimum energy of 3\,GeV for the jets, a minimum
angle of 10\,degrees from either beam for each jet and a minimum
jet/jet invariant mass of 10\,GeV.

The S/B ratio can be improved dramatically by applying additional
$p_\perp$-cuts and, most significantly, by invariant mass cuts.
Figures~\ref{fig:diff-udw-compt150}a and~\ref{fig:diff-udw-compt150}b
show the $p_\perp$-distributions of the jets and the $W$-boson
produced both by a 150\,GeV Higgs signal and by the corresponding
background (see figure~\ref{fig:udW}).
These distributions suggest a transversal momentum cut of
$p_\perp<50\mathop{\textrm{GeV}}$.  In fact, this value can be
understood from looking at the kinematics of a three-body decay of an
on-shell Higgs. There the transversal momentum of particle~1 is
bounded by
\begin{equation}
   p_{\perp,1} \le \frac{\sqrt{(M_H^2 - (m_1 + m_{23})^2)
                               (M_H^2 - (m_1 - m_{23})^2)}}
                        {2M_H}
\end{equation}
where~$m_{23}$ is the invariant mass of the other two particles.  In
the extreme case~$m_{jj}=0$ we
find for example~$p_{\perp,W}\le(M_H^2-M_W^2)/(2M_H)$ which is
$\approx54\mathop{\textrm{GeV}}$ for a 150\,GeV Higgs.

Using a transverse momentum cut of $10
\mathop{\textrm{GeV}}<p_\perp<50\mathop{\textrm{GeV}}$ on either jet
and the $W$-boson, we find the invariant mass distributions for signal
and background, shown in figure~\ref{fig:diff-udw-compt150}c.
Obviously, a cut on the invariant mass of $M_H -
5\mathop{\textrm{GeV}} < M_{\gamma\gamma} < M_H +
5\mathop{\textrm{GeV}}$ brings the signal to the level of the
background.  This cut can be applied because, as mentioned in the
introduction, we assume that the Higgs mass will have been determined
by the time a PLC is commissioned.  However, it should be noted from
table~~\ref{tab:wide-hadr} that the S/B ratio of nearly one for a
Higgs mass at 140\,GeV deteriorates when $M_H$ approaches the $WW$
threshold.

The fully hadronic final state suffers from additional combinatorial
background, since the four jets can not be assigned unambiguously.
Therefore a more detailed analysis using a full
four-fermion calculation will have to be performed in the future.
However, since we assume one of the $W$'s to be on-shell, an invariant
mass cut on one pair of jets will have rejected most of the
combinatorics.

%%%%%%%%%%%%%%%%%%%%%%%%%%%%%%%%%%%%%%%%%%%%%%%%%%%%%%%%%%%%%%%%%%%%%%%%
\subsection{Scenario~II: Narrow-Band PLC}
\label{sec:narrow-band}

Scenario~II assumes the technical feasibility~\cite{telnov} of
producing a narrow photon spectrum around a previously measured Higgs
mass.   For the numerical illustrations we have chosen again Higgs
masses of 140, 150 and 160\,GeV.

\begin{table}
  \begin{center}
    \setlength{\arraycolsep}{1em}
    \renewcommand{\arraystretch}{1.4}
    \begin{tabular}{c|rr|rr|rr}
       & \multicolumn{2}{c|}{$M_H=140\mathop{\textrm{GeV}}$}
       & \multicolumn{2}{c|}{$M_H=150\mathop{\textrm{GeV}}$}
       & \multicolumn{2}{c}{$M_H=160\mathop{\textrm{GeV}}$}\\
                  cuts & signal/fb & BG/fb
                       & signal/fb & BG/fb
                       & signal/fb & BG/fb                 \\\hline
      \multicolumn{2}{c}{$\gamma\gamma\to\bar udW^+$}      \\\hline
             canonical &  221.2           &   122.3
                       &  309.4           &   262.2
                       &  566             &  1097                 \\
                       & (110.6\rlap{)}   &  (179.2\rlap{)}
                       & (154.7\rlap{)}   &  (429.2\rlap{)}
                       & (283\rlap{)}     & (2493\rlap{)}                \\\hline
             $p_\perp>20\mathop{\textrm{GeV}}$
                       &   56.6           &    10.2
                       &  143.4           &    42.8
                       &  422             &   471                 \\
             $m_{Wud} \pm5\mathop{\textrm{GeV}}$
                       &  (28.3\rlap{)}   &   (19.6\rlap{)}
                       &  (71.7\rlap{)}   &   (82.0\rlap{)}
                       & (211\rlap{)}     & (1139\rlap{)}                \\\hline
      \multicolumn{2}{c}{$\gamma\gamma\to\bar\nu\mu^-W^+$} \\\hline
             $p_\perp>20\mathop{\textrm{GeV}}$
                       &   20.0           &     3.2
                       &   50.2           &    25.1
                       &  148.8           &   250                 \\ 
                       &  (10.0\rlap{)}   &    (6.5\rlap{)}
                       &  (25.1\rlap{)}   &   (53.7\rlap{)}
                       &  (74.4\rlap{)}   &  (593\rlap{)}
    \end{tabular}
  \end{center}
  \caption{\label{tab:narrow}%
    Cross sections for $\gamma\gamma\to\bar udW^+$
    and~$\gamma\gamma\to\bar\nu\mu^-W^+$, assuming a narrow-band PLC
    (scenario~II).  The numbers without brackets correspond to fully
    polarized ($(++)$ or~$(--)$) photon beams.  For comparison, the
    numbers corresponding to unpolarized beams are given in brackets.} 
\end{table}

We expect that the narrow photon energy spectrum reduces the
background cross section substantially and increases the signal at the
same time.  This expectation is confirmed by the numerical results
shown in table~\ref{tab:narrow}.  Since the energy is fixed at the
Higgs mass by the experimental conditions, we can now use also
the semileptonic channel, which we had to disregard in scenario~I,
because the incomplete kinematical information did not allow invariant
mass cuts.

In addition, the photons will be highly polarized in scenario~II such
that states with $J_z = 0$ are strongly enhanced~\cite{telnov}.  Higgs
bosons are produced only from photon helicities of $(++)$ and $(--)$
that correspond to states with $J_z = 0$.  For fully polarized photon
beams, the Higgs signal cross section is enriched by a factor of two
(see~\cite{bbc}) whereas the background is suppressed with respect to
the unpolarized case.  Close to the threshold, we can estimate this
suppression factor simply~\cite{ginzburg} by comparing the $W$ pair
production cross section just above the $WW$ threshold in the
polarized ($J_z=0$) and unpolarized case.  The cross sections
read~\cite{sushkov,boudjema}
\begin{subequations}
\begin{align}
  \sigma_{\gamma\gamma \to W^+W^-}^{\text{unpol}}
     &= \frac{\pi \alpha^2}{2 M^2} \beta
          \left [ 22 - 9 \beta^2 + 3 \beta^4 - 
          \frac{3 (1 - \beta^2)^2 (1 + \beta^2)}{2 \beta} 
          \log \left | \frac{1+\beta}{1-\beta} \right |~ 
         \right ] \\
  \sigma_{\gamma\gamma \to W^+W^-}^{J_z = 0}
     &= \frac{\pi \alpha^2}{2 M^2} \beta 
          (1 + 3 \beta^2) (3 + \beta^2)
          \left [ 1 + 
          \frac{1 - \beta^2}{2 \beta} 
          \log \left | \frac{1+\beta}{1-\beta} \right |~ 
          \right ]
\end{align}
\end{subequations}
with the velocity $\beta = \sqrt{1 - 4M^2/s_{\gamma\gamma}}$.  Close
to threshold, these cross sections are proportional to~$\beta$. The
coefficients, $19/2$ in the unpolarized case and $3$ in the polarized
case, result in a suppression of $W$ pair production close to
threshold through photon polarization by $19/6\approx3$.  The explicit
calculation including all tree diagrams and cuts shows a suppression
factor varying from 2.5 to 1.5 as the Higgs mass moves down from the
$WW$ threshold  (cf.~table~\ref{tab:narrow}).

The results of the complete calculation in table~\ref{tab:narrow}
show that the S/B ratio is smaller than estimated in~\cite{ginzburg}.
If only the very inclusive canonical cuts are applied, the S/B ratio
varies from 1.8 for a 140\,GeV Higgs mass to 0.5 for a 160\,GeV Higgs
mass close to the $WW$ threshold.  This is smaller than the values
estimated in~\cite{ginzburg} by factors of 4 and 1.7, respectively.
Of course, if $p_\perp$-cuts on jets, leptons and the $W$-boson are
applied, as well as the invariant mass cut in the case of the hadronic
final state, the S/B ratio improves just like in scenario~I.  However,
we should stress that even for the most stringent cuts, the S/B ratio
approaches only 5.6 for a 140\,GeV Higgs mass, which remains below the
8 estimated in~\cite{ginzburg} \emph{without any cuts}.

In any case, the S/B ratio with optimized cuts is sufficiently large
and approaches unity for the hadronic final state, even for a Higgs
mass of 160\,GeV close to the $WW$ threshold.
  
In addition to the improved S/B ratio, the signal cross sections are
also 10 times larger in the narrow-band scenario~II than in
scenario~I.  Assuming a luminosity of 30\,--\,50\,
$\mathop{\textrm{fb}}^{-1}\mathop{\textrm{year}}^{-1}$~\cite{telnov},
we can expect sufficient statistics at a PLC operating in scenario~II.

%%%%%%%%%%%%%%%%%%%%%%%%%%%%%%%%%%%%%%%%%%%%%%%%%%%%%%%%%%%%%%%%%%%%%%%%
\section{Conclusions}
\label{sec:concl}

We have investigated the options for extracting a signal for an
intermediate mass Higgs in the process $\gamma\gamma\to
W+2\mathop{\textrm{fermions}}$ at a future Photon Linear
Collider~(PLC).  The more realistic signal to background~(S/B) ratios
are significantly smaller than those obtained earlier~\cite{ginzburg}
for $\gamma\gamma\to W^*W$.  The S/B ratio is of course most
favourable for a narrow band PLC with energies tuned to the Higgs
mass.  For a wide band PLC the hadronic final state provides enough
kinematical information to raise the S/B ratio close to unity.

If a PLC can be constructed~\cite{telnov}, it will therefore play an
important part in determining the properties of an intermediate mass
Higgs boson in general and in measuring the $\gamma\gamma H^0$ vertex
in particular.

%%%%%%%%%%%%%%%%%%%%%%%%%%%%%%%%%%%%%%%%%%%%%%%%%%%%%%%%%%%%%%%%%%%%%%%%
\subsection*{Acknowledgements}

We have benefitted from discussions with I.~Ginzburg and V.~Serbo,
in particular concerning the narrow-band PLC and photon polarization.
E.~B., V.~I., and A.~P. are grateful to DESY IfH Zeu\-then for the
kind hospitality, where part of the work has been done and to
P. S\"oding for his interest and support.  The work has been supported
in part by the RFBR grants 96-02-19773a and 96-02-18635a and by the
grant 95-0-6.4-38 of the Center for Natural Sciences of State
Committee for Higher Education in Russia. E.~B. and T.~O. are grateful
to the Deutsche Forschungsgemeinschaft~(DFG) for the financial support
(project MA~676/5-1).

%%%%%%%%%%%%%%%%%%%%%%%%%%%%%%%%%%%%%%%%%%%%%%%%%%%%%%%%%%%%%%%%%%%%%%%%

\end{document}